\documentstyle[12pt,graphicx]{article}
\def\u#1{\verb!#1!\endgroup}

\def\HW{{\small Herwig++}}
\def\FHW{{\small HERWIG}}

\begin{document}
\tolerance=100000
\thispagestyle{empty}
\setcounter{page}{0}
 \begin{flushright}
Cavendish-HEP-06/18\\
CERN-PH-TH/2006-182\\
IFJPAN-IV-2006-6\\
IPPP/06/55\\
KA-TP-10-2006\\
September 2006
\end{flushright}

\begin{center}
{\Large \bf Herwig++ 2.0 Release Note}\\[4mm]

{S. Gieseke\\[0.4mm]
\it Institut f\"ur Theoretische Physik, Universit\"at Karlsruhe\\[0.4mm]
E-mail: \tt{gieseke@particle.uni-karlsruhe.de}}\\[4mm]

{D.\ Grellscheid\\[0.4mm]
\it IPPP, Department of Physics, Durham University\\[0.4mm]
E-mail: \tt{david.grellscheid@durham.ac.uk}}\\[4mm]

{K. \ Hamilton\\[0.4mm]
\it IPPP, Department of Physics, Durham University\\[0.4mm]
E-mail: \tt{keith.hamilton@durham.ac.uk}}\\[4mm]

{A.\ Ribon\\[0.4mm]
\it  PH Department, CERN\\[0.4mm]
E-mail: \tt{Alberto.Ribon@cern.ch}}\\[4mm]

{P.\ Richardson\\[0.4mm]
\it IPPP, Department of Physics, Durham University\\[0.4mm]
E-mail: \tt{Peter.Richardson@durham.ac.uk.ac.uk}}\\[4mm]

{M.H.\ Seymour\\[0.4mm]
\it PH Department, CERN\\[0.4mm]
E-mail: \tt{M.Seymour@rl.ac.uk}}\\[4mm]

{P.\ Stephens\\[0.4mm]
\it Institute of Nuclear Physics, Cracow\\[0.4mm]
E-mail: \tt{stephens@hep.phy.cam.ac.uk}}\\[4mm]

{B.R.\ Webber\\[0.4mm]
\it Cavendish Laboratory, University of Cambridge\\[0.4mm]
E-mail: \tt{webber@hep.phy.cam.ac.uk}}\\[4mm]

\end{center}

\vspace*{\fill}

\begin{abstract}{\small\noindent
    A new release of the Monte Carlo program \HW\ (version 2.0) is now
    available. This is the first version of the program which 
    can be used for hadron-hadron physics and includes the full
    simulation of both initial- and final-state QCD radiation.
}
\end{abstract}

\vspace*{\fill}
\newpage
\tableofcontents
\setcounter{page}{1}

\section{Introduction}

The last major public version (1.0) of \HW\ was reported in detail
in \cite{Gieseke:2003hm}, subsequently a release~(2.0$\beta$)
with minimal hadron-hadron physics was made
available for integration testing within experimental
software frameworks, \cite{Gieseke:2006rr}.
In this note we describe the main modifications
and new features included in the latest public version, 2.0. 
We consider this to be the first version which can be used for realistic 
physics studies of hadron-hadron collisions.

Please refer to \cite{Gieseke:2003hm} and the present paper if
using version 2.0 of the program. A full manual will be released in the near future.

The main new features of this version are that the limited functionality in 
the $\beta$-release for the simulation of hadron-hadron collisions has been
extended to include: the simulation of processes other than Drell-Yan;
final-state radiation of the time-like partons produced in the space-like
shower; initial- and final-state showers in the decay of heavy
particles~(e.g. the top quark); the matrix element correction for Drell-Yan processes
and top quark decays; the simulation of QED radiation in particle
decays using the YFS formalism~\cite{Hamilton:2006xz};
and simulation of the underlying event using the model of the 
UA5~collaboration~\cite{Alner:1986is}.

\subsection{Availability}
The new program, together  with other useful files and information,
can be obtained from the following web site:
\small\begin{quote}\tt
            http://hepforge.cedar.ac.uk/herwig/
\end{quote}\normalsize

\section{Hadron-Hadron Processes}

  A number of matrix elements have been added for important processes in
  hadron collisions 
  including:\footnote{The charge conjugate processes while
	             included in the code have been omitted in the list for brevity.}
\begin{itemize}
\item Direct photon pair production, i.e. $q\bar{q}/gg\to \gamma\gamma$;
\item Photon plus jet production, i.e. $q\bar{q}\to\gamma g$, $qg\to\gamma q$;
\item Higgs boson production, i.e. $q\bar{q}/gg\to h^0$;
\item Higgs boson plus jet production, i.e. $gg\to g h^0$, $qg\to q h^0$ and 
      $q\bar{q}\to g h^0$;
\item Heavy quark pair production, i.e. $q\bar{q}/gg\to Q\bar{Q}$;
\item QCD $2\to2$ scattering processes, i.e. $gg\to gg$, $gg\to q\bar{q}$, 
      $q\bar{q}\to gg$, $qg\to qg$, $qq\to qq$, $q\bar{q}\to q\bar{q}$;
\end{itemize}
 in addition the $W$ and $Z$ Drell-Yan processes which were included in the 
 previous version have been extended to include the hadronic decays of the boson.

 Currently the kinematic reconstruction of the parton showers in these processes
 does not use the original approach advocated in \cite{Gieseke:2003rz} but a
 simpler method where the centre-of-mass energy and rapidity of the hard
 process are preserved in the showering process. This is the procedure
 of \cite{Gieseke:2003rz} for $s$-channel colour-singlet systems.
 The time-like shower of the ongoing partons produced in the initial-state
 shower is generated using the same choice of basis vectors as for the space-like
 shower.

\begin{figure}
\begin{center}
\includegraphics[angle=90,width=0.7\textwidth]{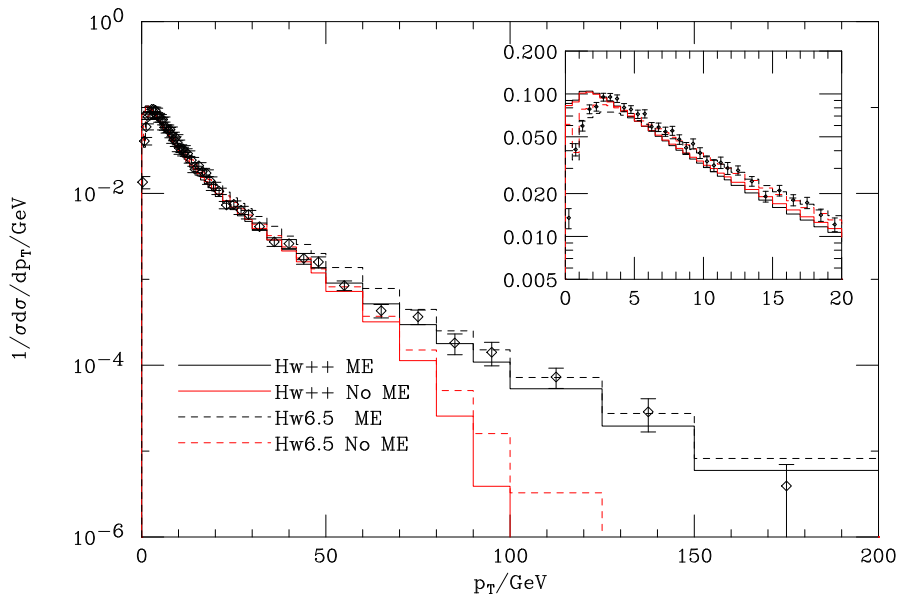}
\end{center}
\caption{The transverse momentum of the $Z^0$ boson at the Tevatron calculated by \FHW6.5
	 and \HW\ compared with run~I CDF data taken from~\cite{Affolder:1999jh} 
         with and without matrix element corrections.}
\label{fig:Zpt}
\end{figure}

 The matrix element correction for the production of an $s$-channel gauge boson,
 either $W^\pm$ or $Z^0$, has been included using the approach of \cite{Corcella:1999gs}.
 The comparison of \HW\ with and without the matrix element correction is shown in
 Fig.\,\ref{fig:Zpt}. \HW\ gives better agreement with the data for medium  
 $p_T$ but peaks about $2$~GeV below data in the low $p_T$ region.
 \HW\ does not yet include a model of intrinsic-$k_T$ in the
 incoming hadrons which could be tuned to reproduce the data.

\section{Top Decay}

  One of the main features of the new shower formalism~\cite{Gieseke:2003rz} was the 
  treatment of the QCD radiation in the decay of heavy particles. This is
  now implemented and includes the initial-state forward shower of the 
  heavy particle in its decay together with the traditional time-like
  shower of the decay products. This is essential in ensuring that the full
  phase-space for the emission of soft gluons is covered, in for example $t\to bW$.

\begin{figure}
\includegraphics[angle=90,width=0.45\textwidth]{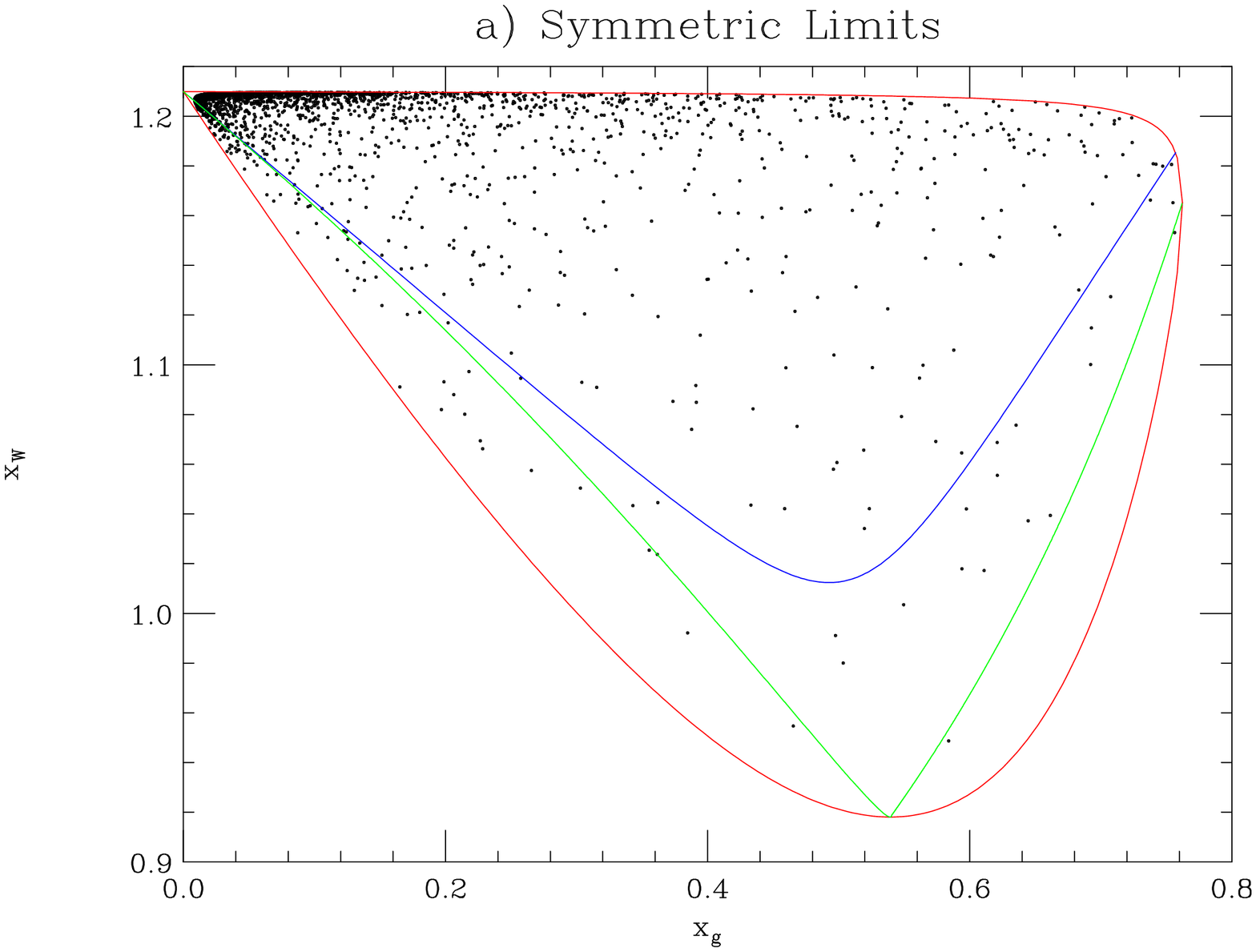}\hfill
\includegraphics[angle=90,width=0.45\textwidth]{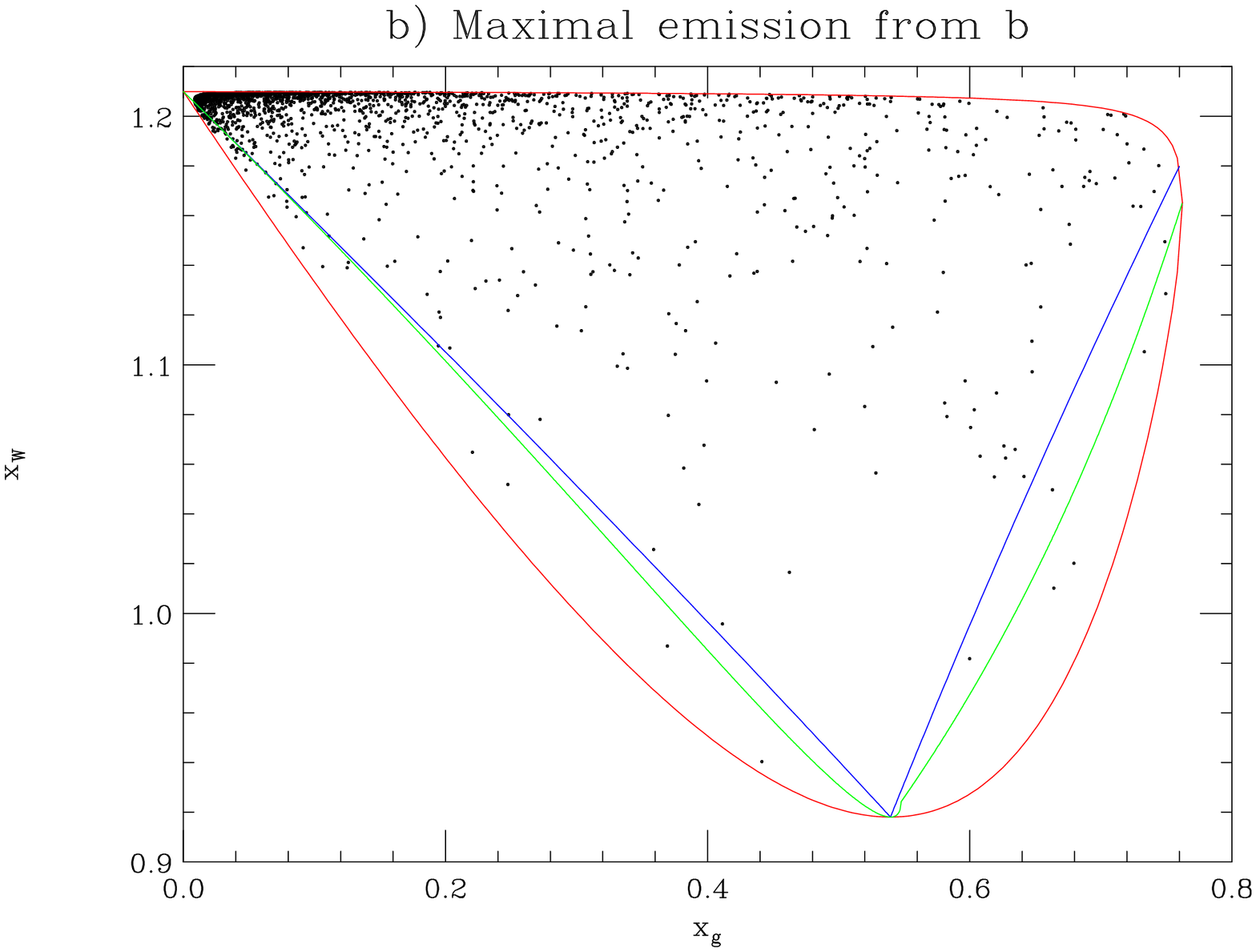}
\caption{Dalitz plot for gluonic radiation in top decay. 
         In both plots both the soft and hard matrix element corrections have
	 been applied, but only one emission has been allowed.
         a) shows the radiation for the symmetric choice of \cite{Gieseke:2003rz}
         for emission from the top and bottom while b) shows the radiation
         with the scales chosen to give the maximum amount of radiation from the 
         bottom quark. The blue~(innermost) line gives the limit for radiation
         from the bottom,
         the green~(middle) line from the top and the red~(outer) line the boundary
         of the phase-space region.}
\label{fig:topdecay}
\end{figure}

  In addition, in order to have the same level of simulation as in the {\sf FORTRAN} 
  program, and to correct for
  the over emission of soft radiation from the outgoing particles, the matrix 
  element correction for $t\to bW$ has been included using the method
  of~\cite{Corcella:1998rs}. This will be described in
  more detail in \cite{Hamilton:2006??} and the forthcoming manual.
  The matrix element correction in \FHW\ was not infrared safe and 
  required an arbitrary cutoff on the gluon energy however as the new algorithm fills
  the whole region for soft emission such a cutoff is not required.
  Fig.\,\ref{fig:topdecay} shows the distribution of the radiation for
  two different choices of the maximal scales for radiation from the top and bottom
  quark.

\section{Underlying Event}

  The simulation of the underlying event currently uses the model of the 
  UA5~collaboration~\cite{Alner:1986is} adapted as described in \cite{hw65} to
  use the cluster model. This is the same as the {\sf FORTRAN} implementation and
  is intended to give a simple model of the underlying event which will
  be replaced by a more sophisticated approach based on multiple scattering
  in the next major release.

\begin{figure}
\includegraphics[width=0.50\textwidth]{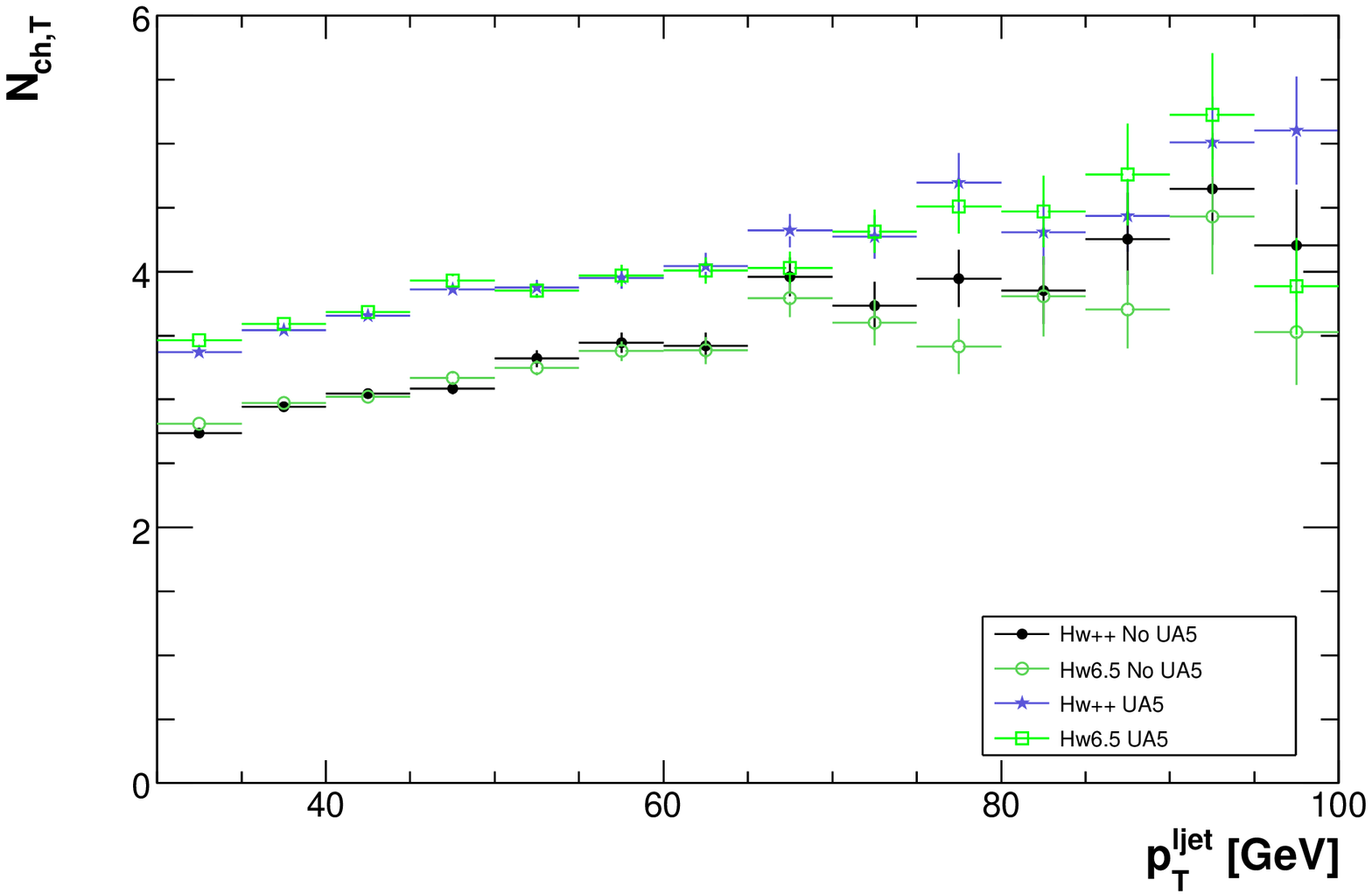}\hfill
\includegraphics[width=0.50\textwidth]{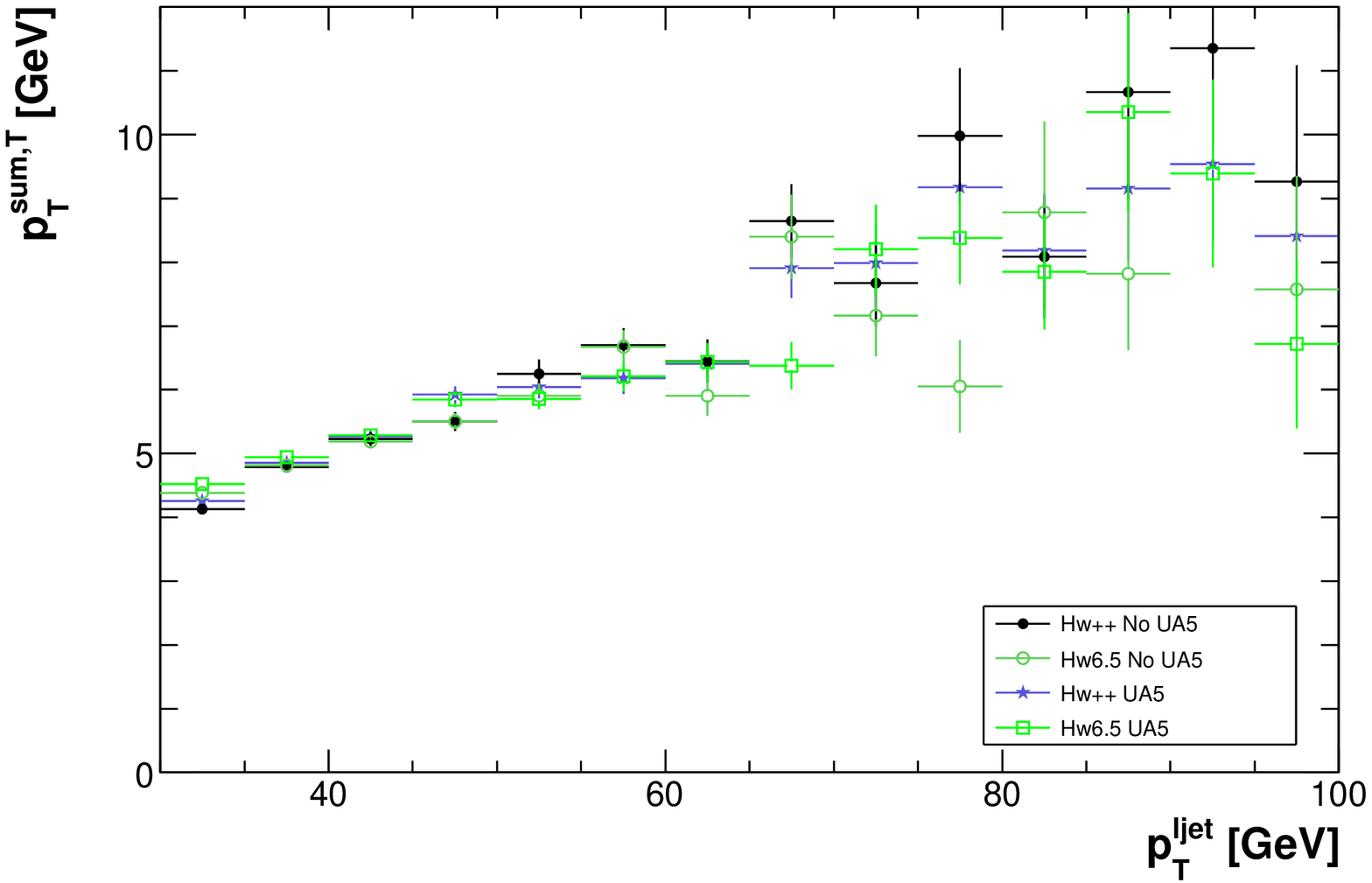}
\caption{The number of charged particles and scalar sum of the transverse momentum
	 in the transverse azimuthal direction with respect to the direction of the 
	leading jet at the Tevatron as a function of the transverse momentum of the 
	leading jet. The blue~(upper) and black~(lower) points
        are the results from Herwig++ with/without the soft underlying event.
        The light green~(upper) and dark green~(lower) points were generated with
 	\FHW\ with/without the soft underlying event. For both the number of charged
	particles and transverse the {\sf FORTRAN} results lie slightly above those
	from \HW. The observables are defined in more detail in~\cite{Affolder:2001xt}.}
\label{fig:ua5}
\end{figure}

  Fig.\,\ref{fig:ua5} shows the number of charged particles and transverse momentum
  transverse to the direction of the leading jet at the Tevatron. In general there is
  good agreement between \HW\ and \FHW\ for the distributions of particles from the 
  soft underlying event with \HW\ producing more particles from the perturbative
  scattering.

  In addition, the simple model of the forced branching after the initial-state parton
  showers, to ensure that the evolution
  ends on the valence quarks in the incoming hadron, has
  been replaced with a more sophisticated approach based on that in the 
  {\sf FORTRAN} program~\cite{hw65}. This model allows the forced branchings to
  take place between a scale {\sf QSpac}~(Default 2.5~GeV) and a multiple 
  {\sf EmissionRange}~(Default~1.1) of the minimum scale. The energy fractions
  are then determined using the perturbative result. This replaces the simple model
  in the beta release which generated the scale at the minimum value and used a 
  flat distribution for the energy fraction which tended to give a large number
  of low mass soft collisions.
 
\section{QED Radiation}

\begin{figure}
\includegraphics[angle=90,width=0.45\textwidth]{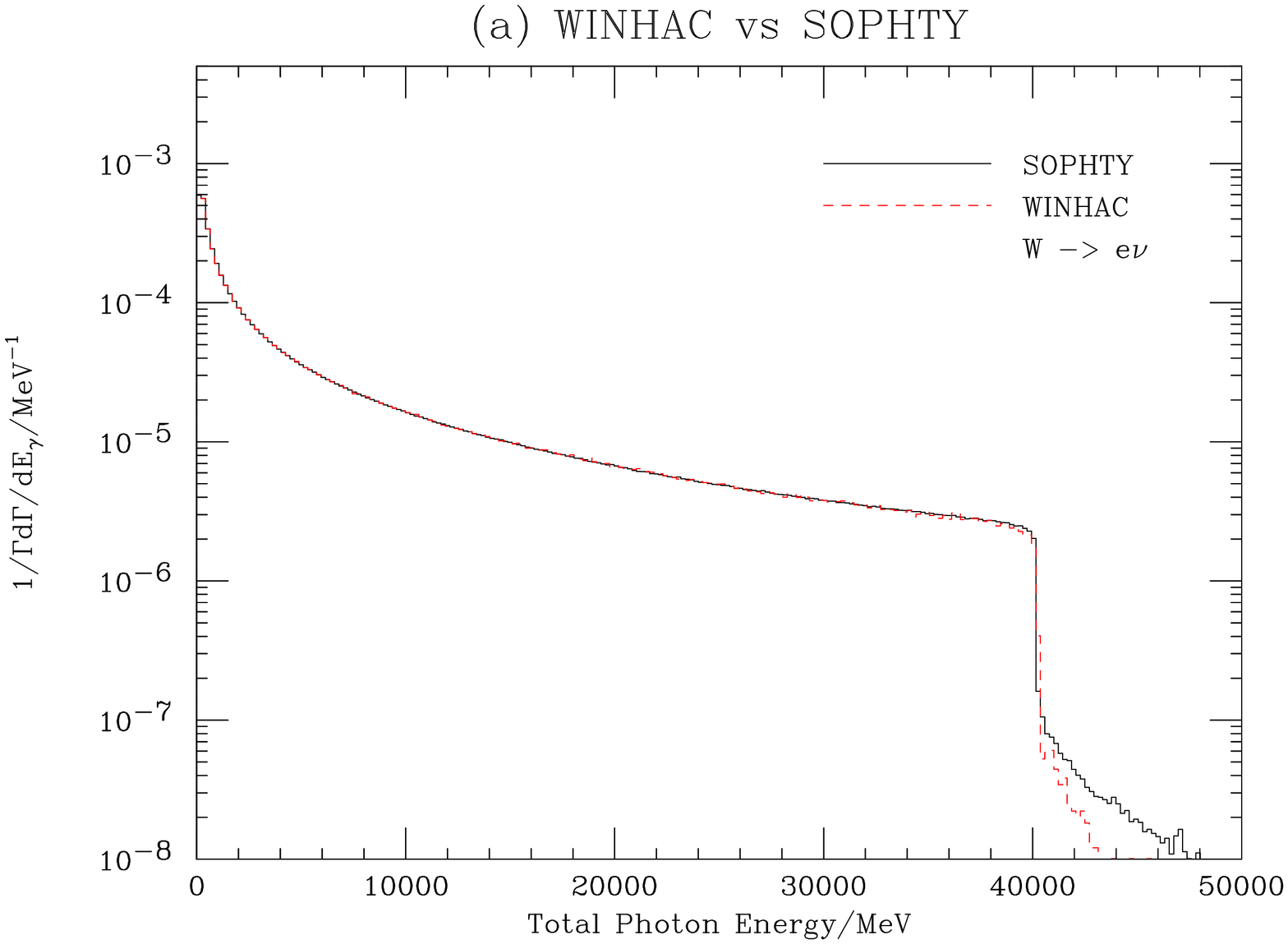}\hfill
\includegraphics[angle=90,width=0.45\textwidth]{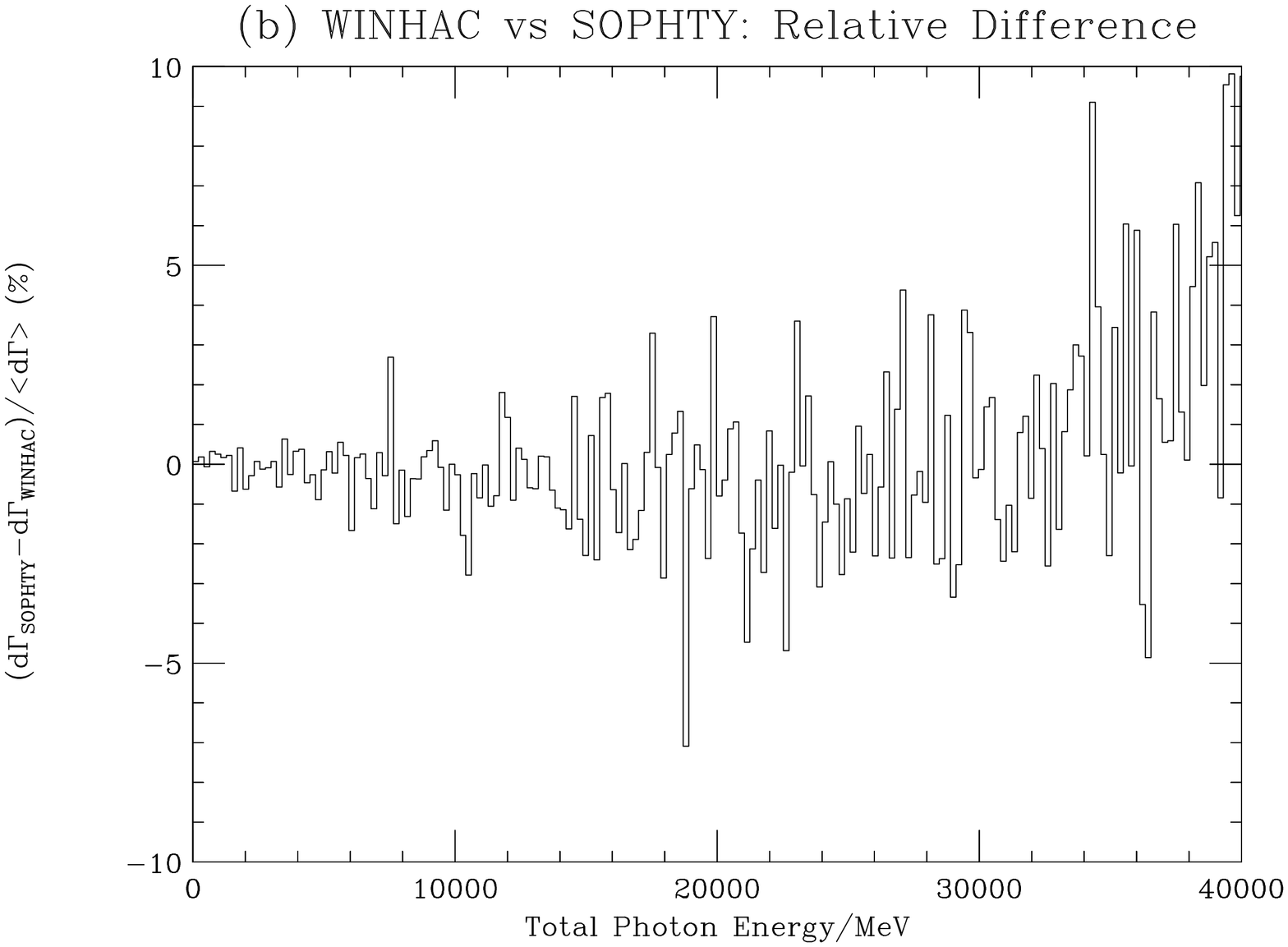}
\caption{The total energy of the photons radiated in 
          ${\rm W}^\pm\to{\rm e}^\pm\nu_e/\bar{\nu}_e$ decays. In figure~(a)
 	the red~(dashed) histogram was generated using the {\sf WINHAC}~\cite{Placzek:2003zg}
        program,
	while the black~(solid) line was generated using {\sf Herwig++}. Figure~(b)
	shows the difference between the spectra shown in~(a) divided by the 
	statistical error. The disagreement about 40~GeV is due to events with at
	least two hard photons which neither program is designed to model well.
	}
\label{fig:YFS}
\end{figure}

  The simulation of QED radiation using the approach of \cite{Hamilton:2006xz}
  has been included for both particle decays and unstable $s$-channel resonances
  produced in the hard process. This approach is based on the YFS
  formalism~\cite{Yennie:1961ad} which takes into account large soft photon
  logarithms to all orders. In addition, the leading collinear logarithms
  are included to $\mathcal{O}\left(\alpha\right)$ by using the
  dipole splitting functions. A comparison of the results 
  of {\sf Herwig++} and  {\sf WINHAC}~\cite{Placzek:2003zg} 
  is shown in Figure\,\ref{fig:YFS} for leptonic W decays, more examples
  and a full description of the approach can be found in~\cite{Hamilton:2006xz}.

  By default this is switched off for both decays
  and hard processes but can easily be switched on by including the 
  {\sf QEDRadiationHandler} 
  as a {\sf PostSubProcessHandler} for the hard process or using the {\sf PhotonGenerator}
  interface of the relevant {\sf Decayer}.

\section{Other Changes}

A number of other more minor changes have been made.
The following changes have been made to improve the physics 
simulation:
\begin{itemize}
\item {\sf Decayers} have been added for top and electroweak gauge boson
      decays which are switched on by default;
\item The default mass of the Higgs boson has been increased and the appropriate
      decay modes added. In addition a number of {\sf Decayers} have been added for
      the various Higgs decays. 
\item Some changes have been made to give a more
      reasonable physical description of the splitting of beam clusters.
\item The default $p_T$ cut for particles in the hard process has 
      changed so that different cuts can be used for different types of particles.
      The default cut for photons and jets~(gluons and quarks other than top)
      has been raised to~$20$~GeV. The cut for top quarks and leptons has been
      reduced to zero;
\item A number of {\sf AnalysisHandlers} for the showering from different processes
      have been added;
\item The {\sf cut} member of {\sf ClusterFissioner} has been made virtual
      so that the model of \cite{Odagiri:2006mc} can be implemented as an external
      package;\footnote{This is available from
                       \verb*1http://hepforge.cedar.ac.uk/herwig/1.}
\item An {\sf AnalysisHandler} to print a tree history of the event using {\sf Graphviz}
      has been added;
\item A new set of input files to run top quark pair production in $e^+e^-$ collisions
      at 500~GeV has been added;
\item The {\sf ShowerAlphaQCD} 
      class implementing strong coupling used in the shower has been improved so that
      the same options as in {\sf FORTRAN} \FHW\ 
      can be used, the default remains as in the previous version.
\item The {\sf AlphaEM} class implementing exactly the same running electromagnetic
      coupling as {\sf FORTRAN} \FHW\ has been included. 
\end{itemize}

The following more technical changes to the code structure have been made:
\begin{itemize}
\item The {\sf Shower} module has undergone a significant redesign in order to 
      include the new features in this version, to make implementing new matrix element
      corrections easier and in preparation for improvements such as the multi-scale
      shower and CKKW procedure which are foreseen in future versions;
\item The {\sf GlobalParameters} object has been removed. The effective gluon mass
      is now taken to be the constituent mass of the gluon from its {\sf ParticleData}
      object. This should be set to zero if the Lund string model is used for the 
      hadronization of the event. 
\item The splitting of the hadronic remnant has been moved to be part of the
      {\sf Hadronization} module;
\item The libraries of {\sf AnalysisHandlers} supplied with the release have been
      restructured to separate those which depend on {\sf KtJet} from those which do not;
\item The original obsolete matrix elements have been deleted;
\item The default number of events outputted to the log file has been reduced to 100 due
      to the larger files created when the UA5 underlying event model is switched on;
\item Numerous improvements have been made to the {\sf DOXYGEN} documentation;
\item Several fixes have been made 
      to correct memory leaks in the initialisation of the event
      generator;
\item The {\sf Amegic} interface which has not been used or 
      tested in some time has been removed from the release;
\item The structure of the default input files has been changed and cleaned up;
\item Most classes have been cleaned up to remove unused member functions;
\item Vector and Tensor meson decay base classes have been removed to speed up
      code as little functionality remained in them.
\end{itemize}

The following bugs have been fixed:
\begin{itemize}
\item A bug affecting the gluon splitting function for time-like radiation;
\item A number of bugs affecting the backward evolution of antiquarks;
\item A bug affecting the azimuthal distribution of the gauge boson in Drell-Yan
      processes;
\item Corrections to particle data tables so that gauge and Higgs boson spins are
      correctly set;
\item A bug in the {\sf VSSVertex} where the off-shell scalar wavefunction
      was not correctly included;
\item A bug to the multi-channel decay phase-space integrator which affected a vertex
      with two off-shell particles, which had not previously been encountered.
\end{itemize}

\section{Summary}

  \HW2.0 is the first version of the \HW\ program with a complete simulation of 
  hadron-hadron physics. We look forward to feedback and input from users, especially
  from the Tevatron and LHC experiments.

  Our next major milestone is the release of version 3.0 which will be at least as
  complete as \FHW\ in all aspects of LHC and linear collider simulation. The major
  new features for this version will be the inclusion of Beyond the Standard Model
  physics, a multiple scattering model for the underlying event and the
  full inclusion of the new hadron decay module. In addition depending on the time
  scale for the release a number of improvements to the parton shower may also be 
  included. Following the release of \HW3.0 we expect that support for the 
  {\sf FORTRAN} program will cease.

\section*{Acknowledgements}

  We would like to thank Manuel B\"{a}hr, Martyn Gigg, Seyi Latunde-Dada,
  Simon Pl\"{a}tzer, Kosuke Odagiri and Alexander Sherstnev for commenting
  on preliminary
  versions of the program and Manuel B\"{a}hr for help in testing the simulation of the
  underlying event and producing the plots of the results of the UA5 model.

\end{document}